\documentclass[sigconf]{acmart}

\AtBeginDocument{%
  \providecommand\BibTeX{{%
    \normalfont B\kern-0.5em{\scshape i\kern-0.25em b}\kern-0.8em\TeX}}}

\setcopyright{acmcopyright}
\copyrightyear{2022}
\acmYear{2022}
\acmDOI{XXXXXXX.XXXXXXX}

\acmConference[SIGGRAPH ’22 Posters]{}{August 08--11,
  2022}{Vancouver, BC, Canada}
\acmPrice{15.00}
\acmISBN{978-1-4503-XXXX-X/18/06}

\usepackage{indentfirst}
\usepackage{tablefootnote}
\usepackage{threeparttable}
\usepackage{multirow}
\begin{document}

\title{Guardian Angel: A Novel Walking Aid for the Visually Impaired}

\author{Ko-Wei Tai}
\affiliation{%
  \institution{National Taiwan University}
  \country{}}
\email{b08902080@csie.ntu.edu.tw}

\author{HuaYen Lee}
\affiliation{%
  \institution{National Taiwan University}
  \country{}}
\email{b08902124@csie.ntu.edu.tw}

\author{Hsin-Huei Chen}
\affiliation{%
  \institution{National Taiwan University}
  \country{}}
\email{amber@cmlab.csie.ntu.edu.tw}

\author{Jeng-Sheng Yeh}
\affiliation{%
  \institution{Ming Chuan University}
  \country{}}
\email{jsyeh@mail.mcu.edu.tw}

\author{Ming Ouhyoung}
\affiliation{%
  \institution{National Taiwan University}
  \country{}}
\email{ming@csie.ntu.edu.tw}


\begin{abstract}
    This work introduces Guardian Angel, an Android App that assists visually impaired people to avoid danger in complex traffic environment. The system, consisting of object detection by pre-trained YOLO model, distance estimation and moving direction estimation, provides information about surrounding vehicles and alarms users of potential danger without expensive special purpose device. With an experiment of 8 subjects, we corroborate that in terms of satisfaction score in pedestrian-crossing experiment with the assistance of our App using a smartphone is better than when without under $99 \%$ confidence level. The time needed to cross a road is shorter on average with the assistance of our system, however, not reaching significant difference by our experiment. The App has been released in Google Play Store, open to the public for free.

\end{abstract}



\maketitle

\section{Introduction}
\begin{figure}[h]
  \centering
  \includegraphics[width=0.8\linewidth]{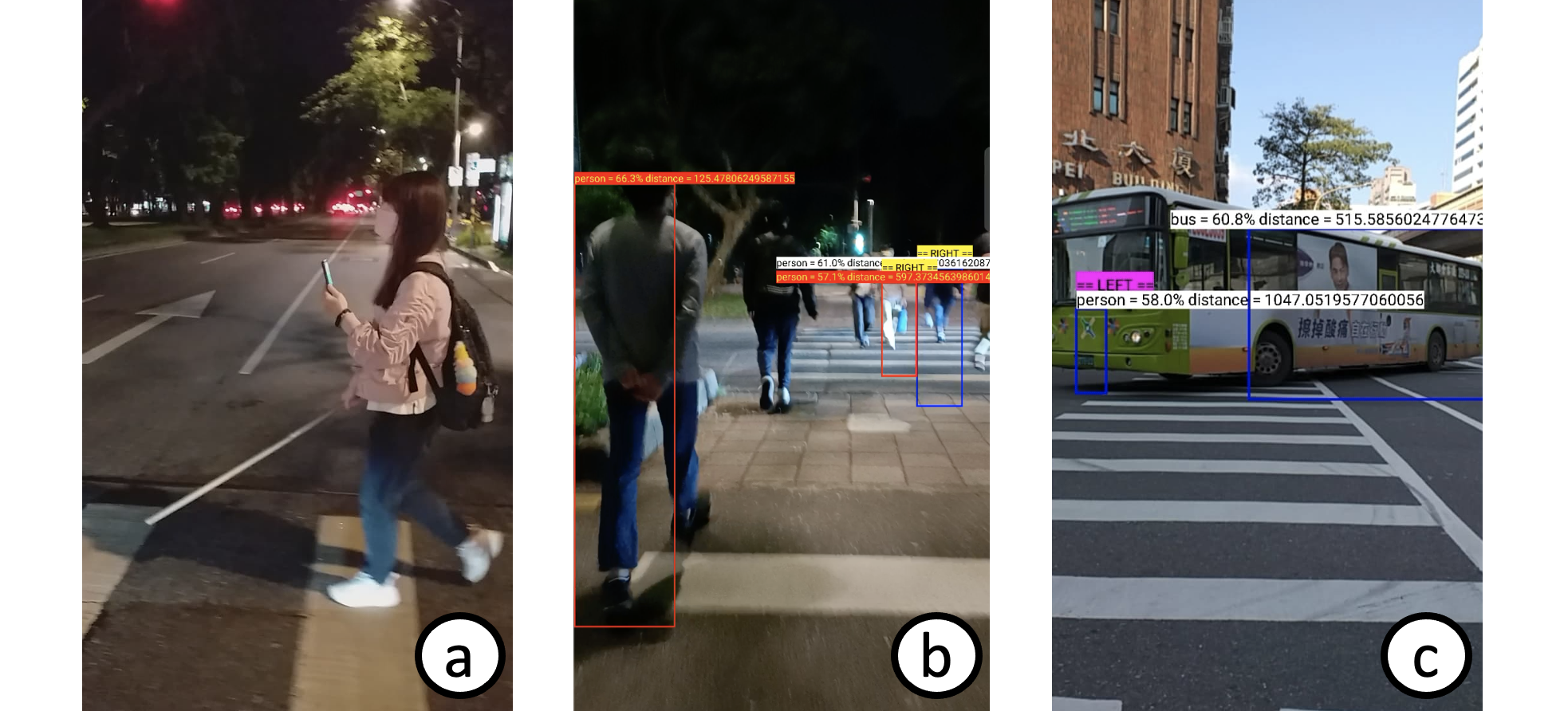}
  \caption{(a) A typical user scenario with both a white stick held in a hand and our App on a smartphone.  (b), and (c), The user interface to show the audio alarm, as well as distance estimation and moving direction.}
  \Description{}
\end{figure}

Crossing a road has been a challenge for the visually impaired people. Even with the help of sound feedback and a white cane, it’s sometimes difficult to identify the direction of a possible danger. There might be a car turning toward people, causing serious danger.\par

There have been several previous works about helping the visually impaired to find objects at home. All of the approaches require special purpose devices, e.g. finger clip, fish eye camera or trackers such as Oculus/HTC trackers.\par
We aim to provide a solution which is easily accessible, affordable, no burden to bring additional special purpose devices to assist visually impaired people on avoiding accidents in a complex traffic environment. In this work, we developed an Android App that uses phone camera to take video from the sight of users. By analyzing the frames, the App is able to distinguish the distance, category of vehicles, and moving direction of objects, then warn the user when some objects might cause danger.

\section{Proposed Approach}

\subsection{Object Detection and Recognition}
We use a pre-trained YOLOv5 model for object detection and recognition. The model was trained on Coco dataset val2017 and deployed to Android by NCNN, a high-performance neural network computing framework optimized for mobile platforms.\par
With its simplicity, YOLO can run more than 45 frames per second when GPU acceleration is applied, making it a great candidate for real-time applications compared to other object detection models, such as R-CNN.

\subsection{Distance Estimation}
Focal length method is applied to estimate the distance between the detected object and camera. For each category of vehicles, we provide the real height by statistics. For example, 140 cm for a car. With the detected object height, we can measure the distance by the equation
\begin{equation}
  \tan \theta = \frac{h}{f} = \frac{H}{D}
\end{equation}
where $h$ is the height in picture, $f$ is focal length
, $H$ is the real height and $D$ is the estimated distance.

\subsection{Moving Direction Estimation}
From interviews with visually impaired users, we realized that knowing which direction an object is coming from enables them to respond to potential danger immediately. Therefore, moving direction estimation is added to the system.\par

First consider the situation when there is only one object detected. As the moving direction in the y-axis (vertical axis of the frame) has been taken care of by distance estimation, we will focus on the moving on x-axis. we compare the x-coordinate of the object in current frame and previous frame, $x_{current} - x_{previous}$, and estimate the moving direction as \emph{right}, \emph{left} or \emph{forward}.\par
By experiments, we discovered that the difference between the detected x-coordinates in two successive frames is not distinct enough to estimate moving direction. Hence, we changed the design to compare current frame and the frame before previous frame. The outcome is significantly improved.

\subsection{Object Matching}
 Now consider the situation when multiple objects are detected at a time. We need to match the objects in successive frames to estimate moving direction. Common object matching methods include intersection of union(IoU) and Euclidean distance, which evaluate closeness using intersection area and Euclidean distance of objects respectively, then match the object with the closest object in the other frame. Although the two methods suffer from ID switching problem in long-time tracking, since we only need to match current object with object in the frame before previous frame, the problem is bearable. For execution efficiency of the App, we abandoned complicated methods such as deep sort and deep learning.\par
 From experiments of IoU method, we found that detected objects in successive frames usually either overlap with almost entire area or don't overlap at all. Thus, Euclidean distance method was chosen for object matching. Note that the matched objects must be detected as the same category.

\subsection{Alarm System}
To utilize the information of object category, distance and moving direction, we set 3 alarm stages, 570 cm to 600 cm, 270 cm to 300 cm, and 120 cm to 150 cm. If any object is within alarm range, the App will warn the user by vibration of 0.8, 1.2, 1.6 seconds long respectively. Meanwhile, we indicate its moving direction by audio clips, e.g. "Car moving left".

\section{Results}

\subsection{Object Detection}
 We tested six pedestrian crossings with different traffic conditions and screen recorded App execution process for analysis. The overall experimental results show that $95.98 \%$ of objects were recognized with the correct category.

\subsection{Moving Direction Estimation}
For a detected object, we compared the estimated moving direction (\emph{right}, \emph{left}, \emph{forward}) with the real moving direction and calculate how many object in different distance ranges are correctly estimated.\par
While average accuracy is $75.39 \%$, the system  achieves $81.24 \%$ accuracy in short and long distance range. The accuracy is comparatively low in middle range because many objects are crowded there. The vibration of users may also affect the estimated direction.

\subsection{User Experience}

\begin{table}
\footnotesize
      \caption{Crossing time and satisfaction scores of users in our experiment}
      \label{}
      \begin{tabular}{ccccccc}
        \toprule
         && \multicolumn{2}{c}{Without our system} && \multicolumn{2}{c}{With our system}\\
        \cmidrule{3-4} \cmidrule{6-7}
        User ID && Time & Satisfaction && Time & Satisfaction\\
        \midrule
        1 && 15.47 & 4 && 13.02 & 5\\
        2 && 21.11 & 1 && 18.55 & 2\\
        3 && 22.54 & 2 && 19.49 & 4\\
        4 && 39.24 & 4 && 28.91 & 4\\
        5 && 18.85 & 2 && 18.08 & 4\\
        6 && 30.13 & 2 && 37.72 & 4\\
        7 && 35.25 & 2 && 24.70 & 5\\
        8 && 16.36 & 4 && 17.22 & 4\\
        \bottomrule
      \end{tabular}
\end{table}

A total of 8 subjects aged between 20 to 66 years old participated in the experiment involving pedestrian crossing of a 10 meters wide road near our campus. Both the satisfaction score and pedestrian-crossing time in seconds were recorded. Two of the subjects are visually impaired. According to the result in Table 1, we corroborate that in terms of satisfaction score with the assistance of our App is better than when without under $99 \%$ confidence level.\par
The time needed to cross the road is shorter on average with the assistance of our system, however, not reaching significant difference by our experiment. Most subjects walked faster and more confidently when our system helped them distinguish between dangers involving cars, pedestrians, or motor-cycles. We observed that subject 6 and 8 took time to think and respond to alarms, so the crossing time is longer when our App is used.

\section{Conclusion and Future Works}

In this work, we propose a method to address the dangers in traffic environment for the visually impaired. With affordable, accessible device, the Android App alarms users when danger is detected to assist them. We've released the App, "Guardian Angel - Walking Aid", on Play Store for free for the visually impaired worldwide.\par

According to the feedback from our interviewees, it will be great if the system not only provide information about coming objects, but also recommend how to avoid the danger. We also plan to filter out some nonurgent alarms, e.g. objects moving away from user, and focus on objects that may cause great danger, e.g. car in a short distance moving toward the user. Furthermore, we wish to provide a direction guidance based on the stripes of a pedestrian crossing.




\begin{thebibliography}{9}
\footnotesize
    \bibitem{texbook} Shih, Meng-Li, et al. "Dlwv2: A deep learning-based wearable vision-system with vibrotactile-feedback for visually impaired people to reach objects." 2018 IEEE/RSJ International Conference on Intelligent Robots and Systems (IROS). IEEE, 2018.
    \bibitem{texbook} Redmon, Joseph, et al. "You only look once: Unified, real-time object detection." Proceedings of the IEEE conference on computer vision and pattern recognition. 2016.
    \bibitem{texbook} Wojke, Nicolai, Alex Bewley, and Dietrich Paulus. "Simple online and realtime tracking with a deep association metric." 2017 IEEE international conference on image processing (ICIP). IEEE, 2017.
   
\end{thebibliography}



\end{document}